\documentclass[aps,prx,onecolumn,floatfix,superscriptaddress,showpacs,amsmath,amssymb,longbibliography]{revtex4-2}
\usepackage{graphicx}
\usepackage{relsize}
\usepackage{hyperref}
\usepackage{amsfonts,amsbsy}
\usepackage{subcaption}
\hypersetup{
    colorlinks = true,
    urlcolor   = blue,
    citecolor  = black,
}

\newcommand{\RomanNumeralCaps}[1]

\usepackage{bm}
\usepackage{color}
\usepackage{epstopdf}
\usepackage{float}
\usepackage{subfloat}
\usepackage[normalem]{ulem}

\DeclareMathAlphabet{\mathsfbi}{OT1}{\sfdefault}{bx}{sl}

\graphicspath{{../fig/}}
\definecolor{C0}{rgb}{0.12,0.47,0.705}
\definecolor{C1}{rgb}{1,0.498,0.055}
\definecolor{C2}{rgb}{0.173,0.55,0.173}
\definecolor{C3}{rgb}{0.839,0.153,0.157}
\definecolor{C4}{rgb}{0.580,0.404,0.741}

\newcommand\Ra{\text{Ra}}
\newcommand\Sc{\text{Sc}}

\newcommand\Ri{\text{Ri}}

\newcommand{\CN}{\mbox{{\it C. augustae} }}

\newcommand{\GL}{{\tt GL}~}
\newcommand{\GE}{{\tt GE}~}

\newcommand\DD{\mathsf{D}}

\newcommand{\be}{\begin{equation}}
\newcommand{\ee}{\end{equation}}
\newcommand{\bea}{\begin{eqnarray}}
\newcommand{\eea}{\end{eqnarray}}

\newcommand{\bx}{{\bm x}}

\newcommand{\Pe}{\text{Pe}\,}

\newcommand{\xa}{{\bm a}}

\newcommand{\xb}{{\bm b}}

\newcommand{\xg}{{\bm g}}

\newcommand{\xj}{{\bm j}}
\newcommand{\xJ}{{\bm J}}

\newcommand{\xSig}{\mathsf \Sigma}

\newcommand{\xu}{{\bm u}}
\newcommand{\xU}{{\bm U}}

\newcommand{\cm}{{\,\rm cm}}

\newcommand{\gm}{{\,\rm gm}}

\newcommand{\mm}{{\,\rm mm}}

\newcommand{\Fig}[1]{Fig.~\ref{#1}}
\newcommand{\subfig}[2]{Fig.~\ref{#1}(#2)}

\begin{document}
\title{Emergent asymmetry in confined bioconvection}

\author{Martin A. Bees}
\affiliation{Department of Mathematics, University of York, York, UK, YO19 5DD.}
\author{Prasad Perlekar}
\affiliation{Tata Institute of Fundamental Research, Hyderabad, Telangana 500046, India.}

\begin{abstract}
Bioconvection is the prototypical active matter system for hydrodynamic instabilities and pattern formation in suspensions of biased swimming microorganisms, particularly at the dilute end of the concentration spectrum where cell-cell interactions typically are neglected.  
Confinement is an inherent characteristic of such systems, including those that are naturally-occurring or industrially-exploited, so it is important to understand the impact of boundaries on the hydrodynamic instabilities.
Despite recent interest in this area we note that commonly-adopted symmetry assumptions in the literature, such as for a vertical channel or pipe, are uncorroborated and potentially unjustified.
Therefore, by employing a combination of analytical and numerical techniques, we investigate whether confinement itself can drive asymmetric plume formation in a suspension of bottom-heavy swimming microorganisms (gyrotactic cells). 
For a class of solutions in a vertical channel we establish the existence of a first integral of motion, and reveal that asymptotic asymmetry is plausible.
Furthermore, numerical simulations from both Lagrangian and Eulerian perspectives demonstrate with remarkable agreement that asymmetric solutions can indeed be more stable than symmetric; asymmetric solutions are in fact dominant for a large, practically-important region of parameter space.
In addition, we verify the presence of blip and varicose instabilities for an experimentally accessible parameter range. 
Finally, we extend our study to a vertical Hele-Shaw geometry to explore whether a simple linear drag approximation can be justified. 
We find that although two-dimensional bioconvective structures and associated bulk properties have some similarities with experimental observations, approximating near wall physics in even the simplest confined systems remains challenging. 
\end{abstract}


\maketitle

\section{Introduction \label{s:intro}} 

In the absence of flows many microorganisms, such as single-celled algae and bacteria, tend to swim in preferred directions, causing accumulations of cells that can drive hydrodynamic instabilities and patterns, termed bioconvection \citep{pk92,hp05,bees2020advances}.
For algae, the biased motion may be upwards against gravity (gravitaxis) or towards light  (phototaxis) both of which can lead to high cell concentrations near the upper surface.
For many cells that are typically more dense than the fluid in which they swim, the resulting stratification may initiate an overturning instability.  
In addition, the cells will also be subject to viscous torques so instabilities can arise in uniform suspensions far from boundaries, called gyrotactic instabilities.  
Gyrotaxis is the result of the combined effect of viscous and gravitational torques, which can cause swimming cells to drift across streamlines and focus in downwelling regions of the flow, with their negative buoyancy driving additional downwelling \citep{phk88}.
Together the two mechanisms of gravitaxis and gyrotaxis are responsible for gyrotactic bioconvection \citep{k85,bees2020advances}, which can occur in a closed container in the absence of other external stimuli (such as chemical gradients, chemotaxis, or light).  

\begin{figure}
\includegraphics[width=0.8\linewidth]{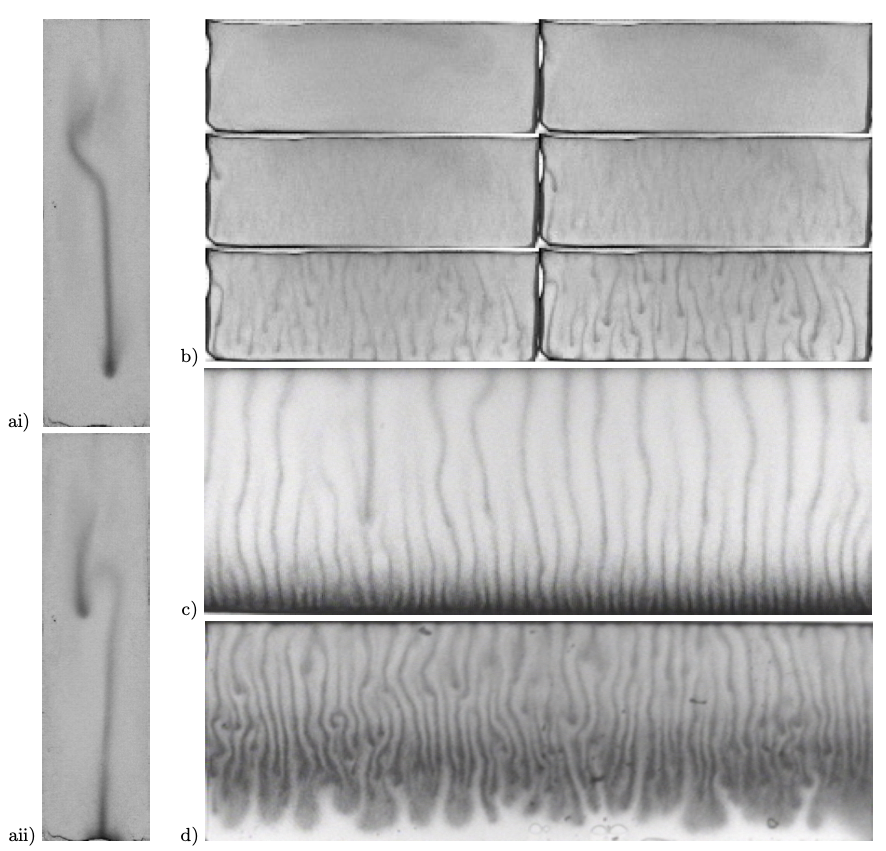}
\caption{\label{fig:plumes}  a) Gyrotactic plumes in a suspension of {\it Chlamydomonas augustae} of concentration $\sim$10$^5$ cells/cm$^3$ illustrating the development of asymmetric gyrotactic plumes in a vertical Hele-Shaw channel between two parallel glass slides of separation 1 mm, and chamber height and width of 7.5 cm and 2.5 cm, respectively, when the mean flux, $Q$, is zero; i) and ii) are 30 s apart. b) \& c) Bioconvection in a suspension of {\it C. augustae} of concentration $\sim 1.6 \times $10$^6$ cells/cm$^3$ in the vertical Hele-Shaw apparatus with a chamber height and width of 2.5 cm and 7.5 cm, respectively, at b) onset (images every 10 s from fully mixed), demonstrating a gyrotactic instability, and c) at a later time (300 s), highlighting the formation of bottom standing plumes. d) Bioconvection in a tapered vertical Hele-Shaw apparatus between two glass slides of width 1 mm at the top and touching at the bottom with cell concentration $\sim 5 \times $10$^6$ cells/cm$^3$; individual gyrotactic plumes are initiated in the top third via a gyrotactic instability, propagate through the middle third and dissipate in the bottom third. Images ai and ii are due to H.~O.~Caldag \& MAB and b-d are due to J.~O.~Kessler \& MAB.}
\end{figure}

Freely swimming microorganisms are ubiquitous in the natural world and impact all aspects of our lives. 
However, many cells spend much of their lives closely associated with boundaries, with biased swimming motion and instabilities occurring in confined environments, from tubes (see Fig.~\ref{fig:plumes}a) to porous structures, such as in soil, and antibiotic resistant biofilm formation on practically every surface from medical devices to ship hulls \cite{sauer2022biofilm}.  
Industrial applications are plagued by inefficiencies due to the fouling of industrial photo-bioreactor walls \cite{borowitzka1999commercial} and we, of course, rely on the proper functioning of our closely confined microbiome. Therefore, it is necessary to gain a proper understanding of suspensions, their swimming behaviour in flows in tubes and channels and their potentially complex interaction with boundaries.
The study of bioconvection in elementary confined environments allows us to develop such knowledge and pursue a systematic process of simplification.

Active matter systems of motile or force-exerting particles, that consume energy to move and are thus out of equilibrium, provide a natural framework for theoretical investigations \citep{simha2002hydrodynamic, pedley10a, ramaswamy2010mechanics}. In dense suspensions, hydrodynamic stresses associated with the swimming motion of individual cells can not be ignored, and the Stokesian instabilities studied in the context of active polar matter become relevant \citep{simha2002hydrodynamic,pedley10a,chat21,rana24}.  
In this article, we shall restrict attention to the non-trivial dynamics of dilute suspensions; the cells swim in a mean direction that is a function of the gradient of the fluid velocity and are advected by the flow that they collectively drive by their negative buoyancy. 
However, we note that there is much to be explored in the intermediate semi-dilute regime, particularly in confined environments where we expect active stresses, such as swimming induced stresslets, to play a qualitative role \cite{pk92,simha2002hydrodynamic,thampi_2022}.

Consider a suspension of algal cells in a downward channel flow. 
At small volume fractions,  the focused cells form plume structures along the centreline \citep{k85,k85a,bees2020advances}. 
An intriguing observation in experiments is the occasional migration of plumes of cells towards the vertical walls, which is typically attributed to a small degree of tilt or to phototactic bias under illuminated conditions (Fig.~\ref{fig:plumes}a), but are there other, hydrodynamic mechanisms for this asymmetry?
 
In Fig.~\ref{fig:plumes}b and c we see bioconvection plume structures forming in a vertical Hele-Shaw apparatus, at early and late times, illustrating both gyrotactic instabilities and long-time bottom standing structures, respectively.  The distance between the two glass plates in this experiment is just 1 mm, and so it is tempting to attempt to average in the direction into the page and thus consider only a two-dimensional bioconvection structure in the horizontal and vertical directions.  But can this be justified?  In fact, there are upwelling regions as well as downwelling, and whilst cells will tend to focus at the centre of the channel where the local flow is downwelling, they will migrate towards slow-moving flow at the boundaries in upwelling zones; cells will be transported at a greater velocity than the cross-sectional mean flow when it is downwards and at a smaller velocity than the mean when it is upwards.  
In addition, it is not clear whether the flow and cell concentration profiles are likely to be symmetric across the channel.  These aspects will play a very interesting role in the overall pattern formation and the resolution will provide the foundation for systematically constructing an averaging process. 

In this paper, we investigate the above questions directly by considering solutions in a two-dimensional vertical channel geometry (see \Fig{fig:sketch}), with some unexpected results.  
There is a surprising complexity of solutions from equations that, at first glance, look relatively straightforward.
We explore the solutions using both analytical and numerical methods, making good use of two distinct simulation techniques, Eulerian and Lagrangian, that generate results with remarkably good agreement.
To access all solutions one must choose the model formulation and its boundary conditions very carefully and in a non-trivial way, not least because it is hard to approximate a solution constrained by no-flux boundary conditions if there are few swimmers present at the boundary.
In addition, the presence of the cells can drive fluid flow down the channel, but so can a pressure gradient. 
These two driving mechanisms combine to determine the mean flux, but it is not unique.  
Alternatively, one may attempt to impose the mean flux to establish the required pressure gradient but, once again, uniqueness is not guaranteed, particularly for small positive or negative flow rates.  
In particular, and in contrast to investigations in the literature \citep[e.g.][]{hwang14a}, we do not impose symmetry (it is easier to enforce zero cell flux in the centre of the channel where cells are present). 
It turns out that asymmetric solutions arise naturally and are more stable in some instances than symmetric solutions.  
We explore the nature of the nonlinear bifurcation structure as the mass flow and pressure are varied and investigate the secondary blip and varicose instabilities using the two distinct numerical approaches.  
This structure depends to a large extent on the non-dimensional parameters, namely the Schmidt, Richardson, and 
Peclet numbers.
This allows us to construct a modified quasi-two-dimensional theory for bioconvection in a Hele-Shaw cell and similar geometries, which for some aspects compares favourably with experimental results.
Finally, we discuss our results in the context of the literature and describe some open questions.

\section{Experimental observations}

Suspensions of biased motile cells often exhibit hydrodynamic instabilities that are reminiscent of classical convection patterns when shallow layers are observed from above \cite{pk92,bees2020advances}. For deeper cultures of cells the emergent structures typically consist of a succession of descending plumes that can form either at the upper surface or, for gyrotactic cells, throughout the suspension.  Thus to explore the initiation of gyrotactic plumes and long term behaviour we conduct experiments in a vertical Hele-Shaw apparatus.  

We follow the general cell culture and quantitative experimental approach of Ref.~\cite{williams11jeb} for \CN with the following adjustments.  Cultures of motile \CN (Wille) cells were grown on Bold's Basal Medium under cool-white lights (1500 lx) with 12 h:12 h L:D photoperiod. Experiments were conducted at a constant temperature of 20$\pm2$~$^{\circ}$C on concentrated cultures (3-5 weeks old) in the middle of their light period when the cells appear consistently active.
Cells were concentrated by allowing the negatively gravitactic cells to swim upwards through absorbent cotton wool placed at the top of the culture flask. 
To avoid the cells sticking to glass surfaces, the Hele-Shaw apparatus was rinsed in a small amount of the culture medium of low cell concentration.
Illumination was provided by a stable 660 nm red LED diffuse backlight array (Edmund Optics), avoiding heating and a phototactic response \cite{foster80}. 
Vessels were constructed from microscope slides (thickness 1.055 mm) bonded together with UV curing glass glue: parallel sided Hele-Shaw cells were constructed from two slides with spacers all around; tapered Hele-Shaw cells were similarly constructed but no spacer was used at the bottom of the container.
Mixing of the suspension was achieved using a metal wire.
Further details of the experimental results will be presented elsewhere.

Fig.~\ref{fig:plumes}b demonstrates a gyrotactic instability, with plume structures appearing in a few tens of seconds throughout the fluid layer.  Over time these individual plumes reorganise into more regularly spaced plumes, as seen in Fig.~\ref{fig:plumes}c.  Plumes initiated at the top of the suspension are associated with an overturning instability (upswimming dense cells accumulate at the top), which interact further down with plumes formed by gyrotactic instabilities. Towards the bottom of the chamber we see ``bottom standing plumes" (BSP) which sit upon a concentrated layer of cells at the bottom.  BSP have been described previously \cite{pk92} but the precise mechanism for their existence is somewhat unclear. Both the BSP phenomenon and the high concentration of cells below the BSP are self-generated and, paradoxically, due to the tendency of bottom-heavy cells to swim upwards: gyrotaxis leads cells to be recirculated in the down-welling flows surrounding plumes that descend from above and become entrained in the plume heads when they interact with the lower boundary.  
Interestingly, the wavelength of the long-term pattern decreases with depth; wavelengths are halved midway down and halved again close to the bottom.

Experiments in a tapered vertical Hele-Shaw apparatus display qualitatively different behaviour, as observed in Fig.~\ref{fig:plumes}d.  Here, gyrotactic plumes are initiated in the upper third of the apparatus, descend through the middle third, and dissipate in the bottom third.  Typically, there are no cells at all near the bottom of the vessel (sometimes the bottom quarter, with the precise exclusion zone depending on mean cell concentration), despite adequate vessel width (for example, 250 $\mu$m compared to a cell diameter of less than 10 $\mu$m).  Near the bottom, plumes are observed to slow and broaden sufficiently for cells to swim out and upwards on average.  In addition, we see a grouping of approximately three individual plumes in broader plume heads creating channels of clear upwelling fluid.  This is clearly a distinct behaviour due to the increasing proximity of the vessel walls with depth and thus increasing resistance to fluid flow.  
There are qualitative features of these experiments that we shall return to in Section~\ref{s:hshaw}.

\section{Modelling gyrotactic bioconvection: two approaches}
\label{s:model}
Consider the flow geometries depicted in \Fig{fig:plumes}.  The suspension velocity, $\xu$, is governed by the Navier-Stokes equations with a buoyancy term that takes account of the difference in density between the gyrotactic cells and the fluid, $\Delta \rho$, such that
\be
\rho \frac{D\xu}{Dt} = - \nabla p_{e} + n n_0 \mathcal{V} \Delta \rho \xg + \nabla \cdot \xSig,
\mbox{~~~~~}\nabla \cdot \xu = 0,
\label{eq:ns}
\ee
where $\xu(\bx,t)$ is the velocity of the suspension, $n_0$ is the average number of cells per unit volume, 
$n(\bx,t)$ is the cell concentration normalized by $n_0$ (such that the average cell concentration is equal to unity),  
$p_{e}(\bx,t)$ is the excess pressure, $\rho \equiv 1 \gm~\cm^{-3}$ 
is the fluid density, $\rho+\Delta \rho$ is the cell density, $\mathcal{V}$ is the volume of an individual cell and $\xg$ is the gravitational acceleration.   The stress tensor $\xSig$ may include terms that account for swimming stresslets or Batchelor stresses. However, here we assume that the suspension is sufficiently dilute that it may be considered a Newtonian fluid with $\nabla \cdot \xSig=\mu \nabla^2 \xu$, where $\mu$ is the fluid viscosity. 

As experiments are over a relatively short time,  the birth and death of an individual is ignored.  We shall see that there is an abundance of complexity in this system so we shall assume for simplicity that the cells are spherical and use either a two-dimensional Lagrangian or an Eulerian description to study the cell density dynamics, as laid out below.

\subsection{Lagrangian description}
The evolution equation for  the centre-of-mass position ${\bm X}_\alpha$  and the orientation $\theta_\alpha$ of $N_p$  gyrotactic organisms in a suspension are given by
\begin{eqnarray}
\begin{aligned}
\frac{d{\bm X}_{\alpha}}{dt} &=  {\bm u}({\bm X}_{\alpha},t) + V_s {\bm p}_\alpha, \\
{d \theta}_\alpha &=  \frac{dt}{2}\left[\omega + \frac{1}{B}\sin(\theta_\alpha)\right] + \sqrt{2 d_r dt} \eta_\alpha(t), 
\end{aligned}
\label{eq:lag}
\end{eqnarray}
where the orientation of cell $\alpha=1,\ldots N_p$ is given by ${\bm p}_\alpha=(\cos(\theta_\alpha),\sin(\theta_\alpha))$, $B$ is the gyrotactic reorientation time scale, $d_r$ is the rotational diffusivity, $\omega$ is the vorticity, and $\eta_\alpha$ is a Gaussian random variable with zero mean and unit standard deviation. From the positions ${X}_\alpha(t)$, we obtain the number density field $n({\bm x},t) \equiv \sum_{\alpha} \delta({\bm x} - {\bm X}_\alpha)$ where the Dirac delta function may be approximated by cosine formula [see Eq. (6.28) in \cite{pes02}]. Together \eqref{eq:ns} and \eqref{eq:lag} describe the coupled hydrodynamics and dynamics of swimmers and will be referred to as the \GL equations.

\subsection{Eulerian description}
The number density field $n$ satisfies a cell conservation equation of the form
\be
\frac {\partial n}{\partial t} =  - \nabla \cdot \xJ , \mbox{~~~~with flux~~} \xJ  = \left[ n \left ( \xu + \xU 
\right) - \DD \cdot \nabla  n \right], \label{eq:cell}
\ee
where the flux terms represent advection by the flow $\xu$, drift relative to the flow with mean swimming velocity $\xU (\bx,t)$, and swimming diffusion with diffusion tensor $\DD(\bx,t)$.  
We use the approximate functions obtained by \cite{bearon12disperse} for the dependence 
of the swimming velocity, $\xU$, and the diffusion tensor, $\DD$, on vorticity. These are based on generalized Taylor dispersion theory for biased swimming cells in a steady linear shear flow \citep{hb02}, such that
\begin{equation}
\xU = V_s \begin{bmatrix} -P(\sigma; \xa^x,\xb^x) \\ -\sigma P(\sigma; \xa^y, \xb^y) \end{bmatrix},~{\rm and}~
{\DD} = \frac{V_s^2}{d_r} 
\begin{bmatrix} 
P(\sigma; \xa^{xx},\xb^{xx}) & \sigma P(\sigma; \xa^{xy} \xb^{xy}) \\
\sigma P(\sigma; \xa^{xy} \xb^{xy}) & P(\sigma; \xa^{yy},\xb^{yy})
\end{bmatrix},
\end{equation}
where $\sigma=\omega/(2 d_r)$ is the non-dimensional vorticity,  and 
\begin{equation}
P(\sigma; \xa,\xb)=\frac{a_0 +a_2 \sigma ^2+a_4\sigma^4}{1+b_2 \sigma^2+b_4 \sigma^4},
\label{eq:P_rat_func}
\end{equation}
with the $\xa$ and $\xb$ coefficients given in  Table~\ref{t:fitparams}.

\begin{table}
\caption{Parameter values for the coefficients $\xa$ and $\xb$ obtained for dimensionless shear strength $\lambda\equiv 1/(2 B d_r)=2.2$, adapted from \citep{bearon12disperse}. Superscipts indicate the components.\label{t:fitparams}}
\begin{center}
\def~{\hphantom{0}}
\begingroup
\setlength{\tabcolsep}{10pt} 
\renewcommand{\arraystretch}{1.5}
\begin{tabular}{lccccc}
\hline
\hline
 & $a_{0}$ &$a_{2}$ & $a_{4}$&$b_{2}$ & $b_{4}$\\
\hline
$\xa^x$		&$5.7 \times 10^{-1}$		&$3.66 \times 10^{-2}$	&$0$			&$1.75 \times 10 ^{-1}$	&$1.25 \times 10^{-2}$\\
$\xa^y$		& $2.05 \times 10^{-1}$ 	&$1.86 \times 10^{-2}$	&$0$			&$1.74  \times10 ^{-1}$	&$1.27 \times 10^{-2}$\\
$ \xa^{xx}$	&$5.00 \times 10^{-2}$	&$1.11 \times 10^{-1}$	&$3.71 \times 10^{-5}$&$1.01 \times  10 ^{-1}$	&$ 1.86 \times 10^{-2}$\\
$ \xa^{yy}$	&$9.30 \times 10^{-2}$	& $1.11 \times 10^{-4}$	&$0$			&$1.19\times 10 ^{-1}$	&$1.63 \times 10^{-4}$\\
$ \xa^{xy}$	&$9.17\times 10^{-2}$	&$1.56\times 10^{-4}$	&$0$			&$2.81 \times  10 ^{-1}$	&$2.62\times 10^{-2}	 $\\
\hline
\hline
\end{tabular}
\endgroup

\end{center}
\end{table}

In the Eulerian framework \eqref{eq:ns} and \eqref{eq:cell}--\eqref{eq:P_rat_func} describe the fluid and cell dynamics. In what follows, we will refer to 
 \eqref{eq:ns} and \eqref{eq:cell}--\eqref{eq:P_rat_func} as the \GE equations.

\section{Bioconvection in a vertical channel} \label{s:channel}
\begin{figure}[!h]
  \includegraphics[width=0.75\linewidth]{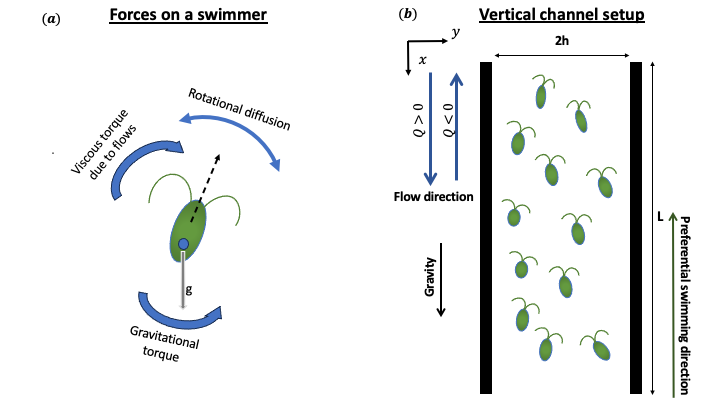}
  \includegraphics[width=0.75\linewidth]{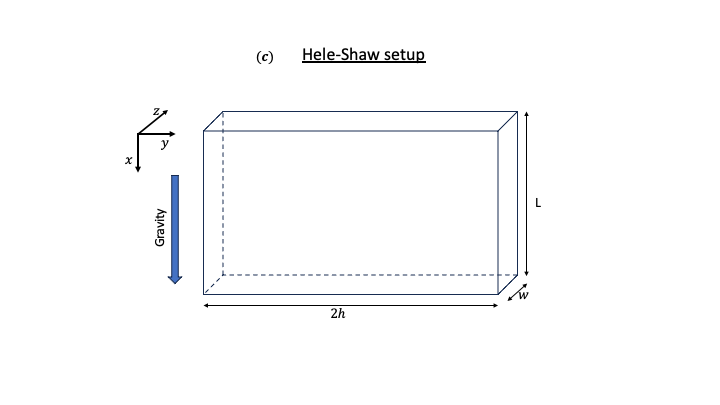}
  \caption{\label{fig:sketch} (a) Sketch showing forces acting on a gyrotactic swimmer, (b) An illustration of a suspension of gyrotactic swimmers in the vertical channel flow geometry considered in section \ref{s:channel}, and (c) the sketch of the Hele-Shaw setup investigated in section \ref{s:hshaw}.}
\end{figure}

The sketch in \Fig{fig:sketch} presents a vertical channel geometry, parallel to the downward pointing $x$-axis with the $y$-axis across the channel.  At the channel walls we impose zero normal cell flux along with no-slip conditions for the flow: $\xj \cdot \xJ (x,0) =\xj \cdot \xJ (x,2h) =0$ and ${\bm u}(x,0)={\bm u}(x,2h)=0$.  To facilitate numerical approximation, we assume that the domain is periodic in $x$ over the length scale $L > h$. Additionally, the flow is maintained at a fixed volumetric flow rate 
\begin{equation}
Q^*\equiv  \int_0^{2h} u_x dy,
\end{equation}
which is, necessarily, constant along the entire channel.

In all of our analysis and simulations, we choose parameters corresponding to a suspension of {\it Chlamydomonas augustae} (a synonym of {\it Chloromonas augustae}, and sometimes identified as {\it Chlamydomonas nivalis}) in a channel of half-width $h=1 \cm$, $\nu=0.01 \cm^2$ ${\rm s}^{-1}$, $V_s=0.0063 \cm$ ${\rm s}^{-1}$,  ${\mathcal V}=5 \times 10^{-10} \cm^3$, $\rho=1 \gm \cm^{-3}$, $\rho + \Delta \rho=1.05\gm \cm^{-3}$, $g=980 \cm$ ${\rm s}^{-2}$, $d_r=0.067$ ${\rm s}^{-1}$, and $n_0 \approx 10^{5}$.

\subsection{Non-dimensional quantities \label{sec:dim}}
By choosing the characteristic length scale as $h$ and the characteristic velocity scale as $V_s$, we find that the dimensionless numbers that describe the flow are the dimensionless flow rate $Q\equiv Q^*/(V_s h)$, the Schmidt number $\Sc\equiv \nu d_r/V_s^2$, the swimming Peclet number $\Pe\equiv h d_r/V_s$, and the Richardson number $\Ri\equiv n_0 {\mathcal V} \Delta \rho g h/\rho V_s^2$. Alternatively, if we choose the characteristic velocity scale as $\nu/h$, we get $Q_1\equiv Q^*/\rho \nu$, $\Sc\equiv \nu d_r/V_s^2$, and the Rayleigh number $\Ra \equiv n_0 {\mathcal V} \Delta \rho g h^3/\rho \nu^2$ as the appropriate dimensionless group. The two dimensionless groups are related  $\Ra=(\Pe/\Sc)^2 \Ri$ and $Q_1\equiv (\Pe/\Sc) Q$.  Following \citep{hwang14}, we characterize the system using $(Q,\Sc, \Pe, \Ri)$.

For the {\it C. augustae} suspension, we choose $\Ri \approx 90$, $\Sc \approx 16.8$, and $\Pe \approx 10.6$. We investigate the suspension dynamics by varying the dimensionless flow rate $Q$. To understand the dependence of our results  on the solution viscosity, for a few cases, we also study the dependence on $\Sc \approx 16.8-80$.

\subsection{Streamwise homogeneous solutions (1D) \label{sec:hs}}
Using the \GE equations, we first investigate the steady state streamwise homogeneous solutions in a vertical channel. Using the incompressibility condition and no-slip boundary condition, the equations for streamwise velocity $u_x(y)$ and number density $n(y)$ become
\begin{equation}
\begin{aligned}
\frac{\partial u_x}{\partial t} &= \nu \frac{\partial^2 u_x}{\partial y^2} + \alpha (n-1) - \frac{dp_e}{dx} ,\\ 
\frac{\partial n}{\partial t} &= \frac{\partial}{\partial y} \left[ -n U_y  + D_{yy} \frac{\partial n}{\partial y} \right],
\end{aligned}
\label{eq:1d}  
\end{equation}
with $\alpha\equiv n_0  {\mathcal V} \delta \rho g h$.

At steady state, we require
\begin{equation}
\begin{aligned}
\nu \frac{d^2 u_x}{dy^2} &=   \frac{dp_e}{dx} - \alpha (n-1),~\rm{and} \\ 
\frac{d n}{dy} 
&=  \left(\frac{a^y_0}{2 V_s a_{0}^{yy}}\right)  \frac{du_x}{dy} n,
\end{aligned}
\label{eq:1dss}  
\end{equation}
where we have assumed weak flow to retain only the leading order terms in $U_y$, and  $D_{yy}$, and imposed the no-flux condition for the number density. Multiplying the first equation by the velocity gradient to rewrite the number density in terms of the velocity derivatives, integrating and re-substituting yields the second-order differential equation
\begin{eqnarray}
\frac{d^2 u_x}{dy^2} - \frac{\gamma}{2} \left(\frac{du_x}{dy}\right)^2 + \beta u_x = \frac{\beta}{\gamma} -A\gamma,
\label{eq:sode1}
\end{eqnarray}
where $A$ is a constant and we define $\gamma\equiv {a^y_0}/{(2 V_s  a_{0}^{yy})}$ and $\beta \equiv (\alpha \gamma/\nu) [1 + (dp_e/dx)/\alpha]$.
Substituting $v = u_x-C/\beta$, where $C \equiv \beta/\gamma - A\gamma$, yields
\begin{eqnarray}
		\frac{d^2 v}{dy^2} - \frac{\gamma}{2} \left(\frac{dv}{dy}\right)^2 + \beta v = 0.
	\label{eq:sode}
\end{eqnarray}
A relationship between $Q$, the pressure gradient, and biological parameters can be found by integrating \eqref{eq:sode1} over the entire domain (using \ref{eq:1dss}), such that
\begin{equation}
C=\frac{\beta}{2h} Q - \frac{\gamma}{4 h} \int_{0}^{2h} \left(\frac{du_x}{dy}\right)^2 dy +  \frac{1}{\nu} \frac{dp_e}{dx}.
\label{eq:CC}
\end{equation}

Equation \eqref{eq:sode} is a differential equation of quadratic Li\'enard-type \citep{strog18}, and following the analysis in \cite{chi87} (alternatively, solving the second of \eqref{eq:1dss} and substituting into the first and comparing with the earlier result) we find the integral
\begin{equation}
\frac{1}{2} \left(\frac{dv}{dy}\right)^2 - \frac{\beta}{\gamma} v - \frac{\beta}{\gamma^2} - k \exp(\gamma v) =0,
\label{eq:integ}
\end{equation}
where $k$ is a constant. It immediately follows that the solutions of \eqref{eq:sode} are a set of periodic orbits surrounding the centre $(v,dv/dy)=(0,0)$, which are bounded by the separatrix defined by $k=0$.

Due to the no-slip boundary condition, we have $u_x=0$ at the walls.  Using \eqref{eq:integ} it is clear that permitted solutions must satisfy $\left.du_x/dy\right|_{y=0} = \pm \left.du_x/dy\right|_{y=2h}$. Note that for the positive case, $dp_e/dx=0$.

\subsubsection{Direct numerical simulations}
Using numerical simulations, we explore the variety of homogeneous solutions and verify the prediction of the integral \eqref{eq:integ}. The spatial domain of length $2h$ is discretized using $N_x=128$ uniformly spaced collocation points. We numerically integrate \eqref{eq:1d} in the vorticity-streamfunction formulation \citep{Perlekar_2009} to maintain a constant flow rate and satisfy the incompressibility condition. The concentration nodes lie between the link joining adjacent vorticity nodes \citep{ghorai99}. Note that in the one-dimensional setting, we may rewrite the vorticity as $\omega=-\partial_y u_x$ and the streamfunction as $\psi=\partial_{yy} \omega$. We run our simulations for at least $t V_s/h > 6 \times 10^5$ time units to achieve a steady state. Unless specified otherwise, we fix $\Ri=90$ and $\Sc=17$.

\subsubsection{Zero mass flow: $Q=0$ \label{sec:zeroq}}
\begin{figure}[!h]
  \centering
  \includegraphics[width=0.4\linewidth]{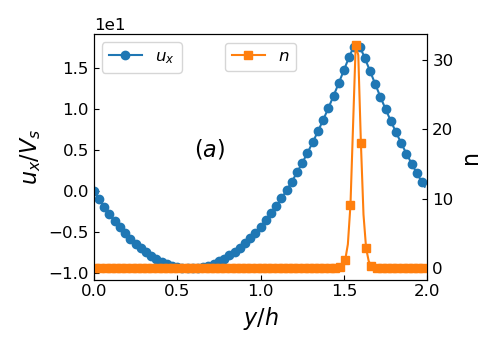}
  \includegraphics[width=0.4\linewidth]{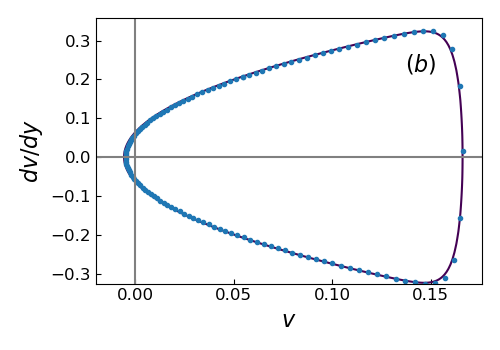}
  \caption{\label{fig:fig1}(a) Plot of the velocity  $u_x$ and the number density $n$, and (b) phase space plot 
  obtained from the DNS (blue dots) and its agreement with the integral curve \eqref{eq:integ} (black line).}
\end{figure}

For $Q=0$, the trivial solution ${\bm u}={\bf 0}$ and $n=1$ satisfies \eqref{eq:1d}. However, this solution is linearly unstable \citep[as noted in][]{pedley10a}. In fact, starting from the trivial solution with random perturbation as initial condition and numerically integrating \eqref{eq:1d} we find that this leads to asymmetric steady state flow and concentration profiles that consist of a near wall down-flow ($u_x>0$) region where, consistent with the gyrotactic mechanism \citep{bees2020advances}, organisms cluster and a single up-flow ($u_x<0$) region that is devoid of organisms (see \subfig{fig:fig1}{a}).

From the velocity field, and its gradients (note $dp_e/dx=0$), we evaluate $C$ using \eqref{eq:CC} and $k$ is obtained by performing a least squares fit to the data. The plot in \subfig{fig:fig1}{b} demonstrates that the points $(v,dv/dy)$ obtained from the DNS are in excellent agreement with the integral curve \eqref{eq:integ}.

\begin{figure}
  \centering
  \includegraphics[width=0.4\linewidth]{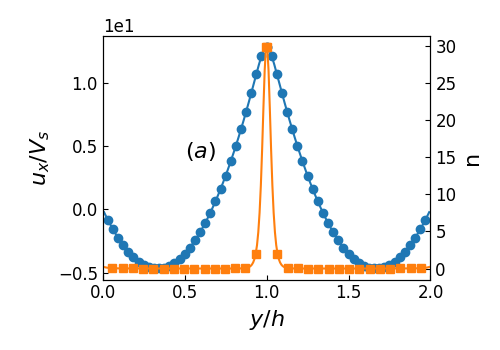}
  \includegraphics[width=0.4\linewidth]{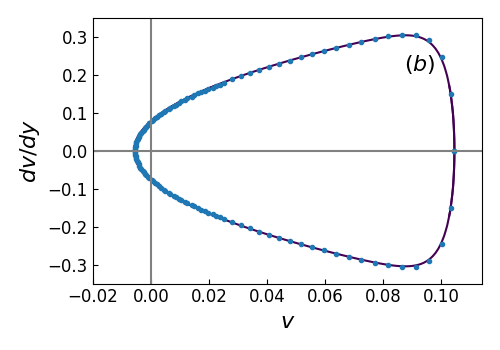}
  \includegraphics[width=0.4\linewidth]{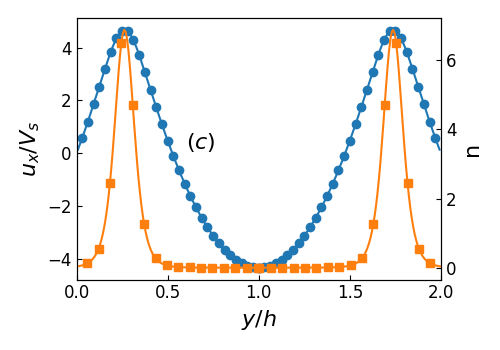}
  \includegraphics[width=0.4\linewidth]{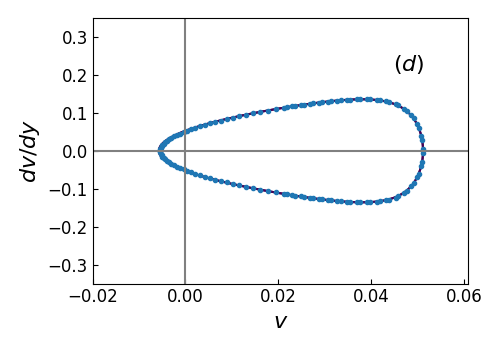}
  \caption{\label{fig:symm}(a,c) Plot of the velocity  $u_x$ (blue circle) and the number density $n$ (orange square) fields for the two symmetric solutions, and (b,d) corresponding phase space plots  
  obtained from the DNS (blue dots) and its agreement with the integral curve \eqref{eq:integ} (black line).}
\end{figure}

{\it Symmetric solutions}--
We identify two more solutions of \eqref{eq:1d} with even symmetry about the centerline (i.e., $u_x(y)=u_x(2h-y)$ and $n(y)=n(2h-y)$). The numerical solutions are obtained by integrating \eqref{eq:1d} over the half-domain and imposing stress-free boundary condition $\omega(h)=0$ about the centreline. The plots in \Fig{fig:symm} show the velocity and concentration profiles. The solutions consist of either a downwelling flow ($u_x>0$) or an upwelling flow ($u_x<0$) near the centreline. We observe $dp_e/dx>(<)0$ in the former (latter) case, and the organisms are concentrated either along the centre or near the two walls. Furthermore, similar to the previous section, the points in the $(v,dv/dy)$ phase plane are in excellent agreement with the integral curve \eqref{eq:integ}. 

\subsubsection{Varying mass flow rate: $Q$ \label{sec:pq}}
\begin{figure}[!h]
  \centering
  \includegraphics[width=0.45\linewidth]{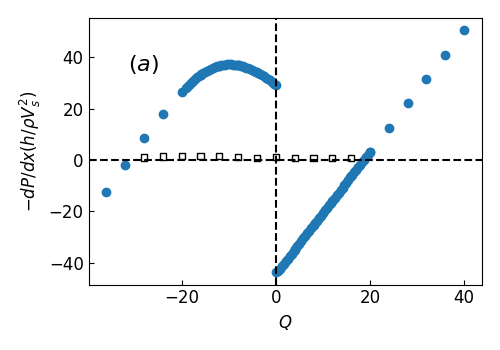}
  \includegraphics[width=0.45\linewidth]{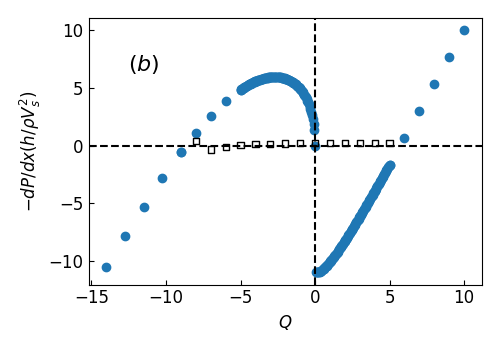}
  \includegraphics[width=0.45\linewidth]{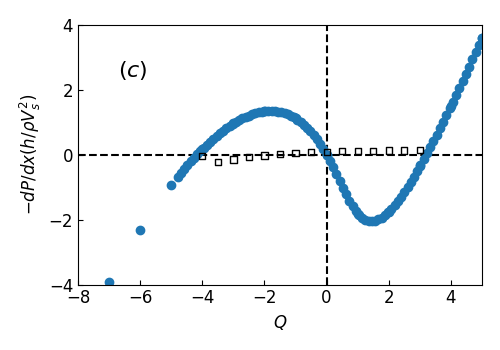}
  \includegraphics[width=0.45\linewidth]{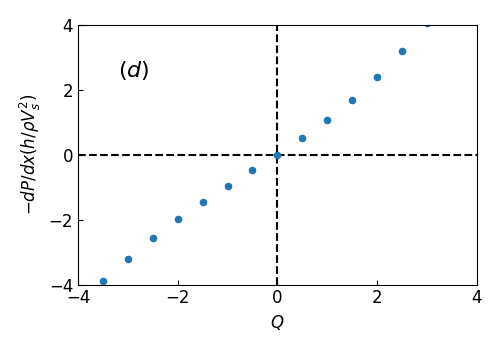}
  \caption{\label{fig:dpdxQ} Plot of $-dp_e/dx$ versus $Q$ for $\Sc=17$ (a), $\Sc=53$ (b), $\Sc=80$ (c), and $\Sc=150$ (d). The asymmetric flow solutions for which $dp_e/dx=0$ are shown with black squares, whereas the symmetric solutions are marked with blue circles.}
\end{figure}

In the absence of organisms, for a Newtonian fluid, the flow rate is linearly related to an excess pressure $Q=-(dp_e/dx)/(12 \mu)$. However, as we observed in the previous section, even for $Q=0$, non-trivial solutions with  $dp_e/dx\neq0$ are possible. Therefore, in \Fig{fig:dpdxQ}, we now characterize the excess pressure by varying $Q$ for different values of $\Sc$.  
For a range of $\Sc$, an intermediate $|Q|$ range exists (about $Q=0$) where the relationship between $dp_e/dx$ and $Q$ is inconsistent with the Newtonian behaviour. 

\begin{figure}[!ht]
  \centering
  \includegraphics[width=0.32\linewidth]{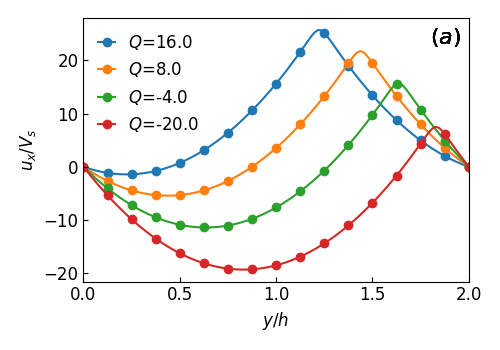}
  \includegraphics[width=0.32\linewidth]{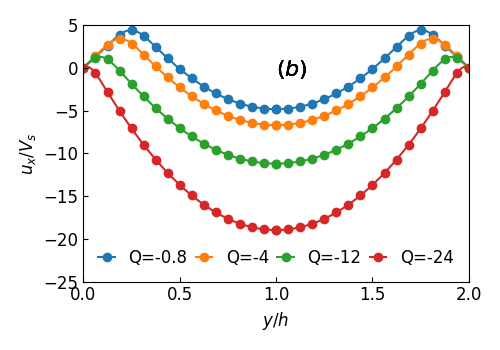}
  \includegraphics[width=0.32\linewidth]{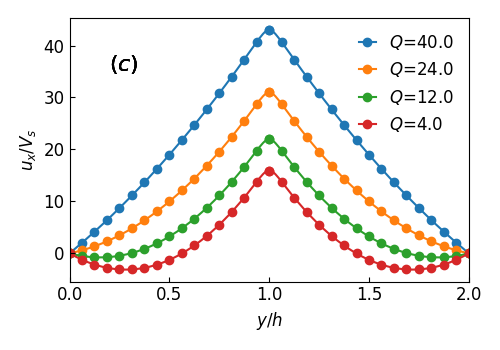}
  \caption{\label{fig:flowQ} (a) The asymmetric flow solution which satisfies $dp_e/dx=0$ for different values of $Q$ at $\Sc=17$. (b,c) 
  Plot of the symmetric velocity profile $u_x$ at $\Sc=17$ and for different values of $Q<0$ (b)  and $Q>0$ (c).}
\end{figure}

{\it Asymmetric solutions} -- The flow profiles for solutions with $dp_e/dx=0$ are asymmetric, as shown in \subfig{fig:flowQ}{a}. For positive $Q$, a downwelling flow region (where $u_x>0$) is present in the vicinity of either one of the walls. This region migrates towards the wall as $Q$ is reduced. As $Q$ becomes negative, an upwelling ($u_x<0$) region also appears. Similarly to the $Q=0$ case, we observe (not shown) that the cells accumulate in regions of downwelling flow. This intermediate $Q$ region disappears for very large $\Sc=150$, and we recover the Newtonian-like monotonic relationship between the excess pressure and the flow rate.

{\it Symmetric solutions} -- In the first and third quadrant of plots in \Fig{fig:flowQ} [i.e.~$(Q>0,-dp_e/dx>0)$ and $(Q<0,-dp_e/dx<0)$] only symmetric solution exists. To obtain the symmetric solution in the other two quadrants, we impose additional symmetry conditions, as in the previous section \ref{sec:zeroq}.  Note that ${\bf u}={\bf 0}$ and $n=1$ is the only symmetric solution that satisfies  $dp_e/dx=0$ at $Q=0$. Therefore, from \Fig{fig:dpdxQ}, it is apparent that the non-trivial symmetric solution with $-dp_e/dx>0$ vanishes around $\Sc\geq 53$, and the solution with $-dp_e/dx<0$ vanishes beyond $\Sc\geq 80$. The specific case for the variation of $dp_e/dx$ with positive $Q$ for $\Sc=80$ agrees with the results of \cite{hwang14a}.
 
The plot in \subfig{fig:flowQ}{b,c} provides the velocity profile for the symmetric solutions for different values of $Q$ with $\Sc=17$. In a similar manner to the $Q=0$ case, we observe that for small positive (negative) $Q$ an upwelling (downwelling) flow is generated near the center. The concentration profile (not shown) has a peak in the downwelling region.  As the qualitative trend of the symmetric solutions for different values of $\Sc$ remains the same, we do not repeat the discussion. The dynamics in the $(v,dv/dy)$ plane also satisfy the integral \eqref{eq:integ}. 

\subsubsection{Comparison  between \GL and \GE approaches}

In the preceding sections, we have used the \GE equations and its simplifications to investigate the dynamics of gyrotactic suspensions. To derive the \GE equations, \cite{bearon12disperse} used generalized Taylor dispersion theory to evaluate the mean biased motion and effective diffusivity in the suspension for a linear shear flow, but it is unclear whether this is a good approximation for flows locally driven by buoyancy. 
Therefore, to further validate our results and ahead of the main discussion of two-dimensional flows below, we now make a comparison of the flow, $u_x$, and cell number density, $n$, profiles obtained by integrating the \GE equations with those obtained by integrating the \GL equations for $Q=0$. Note that the \GL equations are stochastic and explicitly contain the orientation fluctuations. Thus the steady-state profile is obtained statistically by averaging steady states.  For the \GL equations, all parameters are kept identical to the \GE equations, and we use $N_p=4000$ particles.

\begin{figure}[tb]
  \centering
 \includegraphics[width=0.9\linewidth]{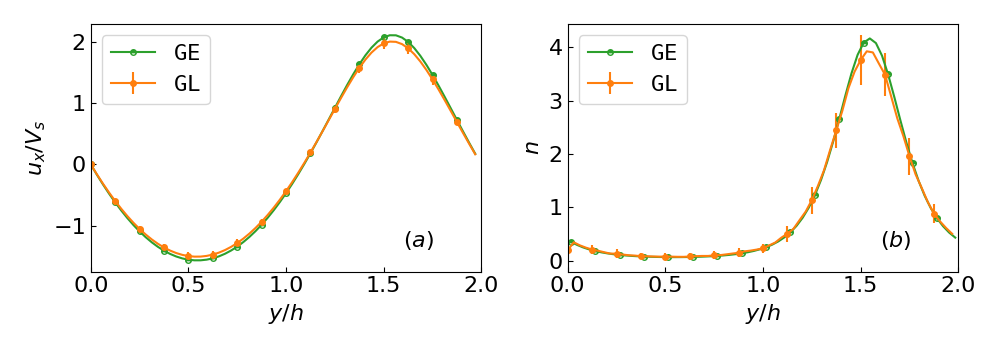}
 \caption{\label{fig:ELcomp} Comparison of the velocity  $u_x$ (a), and the number density $n$ (b) profile obtained by numerically integrating the \GL and \GE equations. The vertical bar denotes the $2 \sigma$ variation  about the average profile obtained from the \GL equations.}
\end{figure}

The plot in \subfig{fig:ELcomp}{a-b} demonstrates good agreement between the profiles obtained by using the two methods. 
These results 
provide confidence in the use of both \GL and \GE approaches in regimes where the organisms modify the flow via buoyancy.

\subsection{Two-dimensional (2D) channel}
The results of our previous section provide good evidence that gyrotactic microorganisms can migrate close to the wall for small flow rates. In particular, around $Q=0$ multiple solution branches can exist, depending on the value of $\Sc$. We now consider a periodic channel flow to understand the dynamics in a spatially extended system as sketched in \Fig{fig:sketch}. 

\subsubsection{Direct numerical simulations}
The channel $(2h,L)$ is discretized using $(N_x,N_y)$  uniformly spaced collocation points. The Navier-Stokes equations are evolved using the vorticity-streamfunction formulation \citep{Perlekar_2009}. In a similar manner to \cite{ghorai99}, the grid is chosen so that the concentration nodes lie in the interior only whereas those of the stream function and vorticity lie in the interior and on the boundary of the domain. All spatial derivatives are evaluated using a second-order finite-difference scheme, and the time-marching in \GE equations is performed using a fourth-order Runge-Kutta scheme. For the \GL equations, we perform the time update using a Euler-Maruyama scheme. The velocity at particle positions and the number density at cell centers using particle locations are obtained using a linear interpolation scheme.

The dimensionless numbers are the same as in Section~\ref{sec:dim}. The channel dimensions $(2h,L)\equiv (2,8)$ are discretised with $(N_x,N_y)\equiv(256,64)$ points. For our \GL simulations, we choose at least $N_p=4 N_x  N_y$ particles, We keep $\Ri=90$ fixed and study the dynamics for different flow rates, $Q$, and the Schmidt number, $\Sc$.

For the $\GE$ simulations we initialize the flow and cell number density by setting 
\begin{equation}
\begin{aligned}
\omega({\bm x},0)=10^{-3} \cos\left[\frac{\pi}{h} (y- h)\right]~~{\rm and~~} 
n({\bm x},0) = 1 + 10^{-2} \sin\left[\frac{\pi}{h} (y- h)\right].
\nonumber
\end{aligned}
\end{equation}

For the $\GL$ simulations, the initial particle orientations, $\theta_{\alpha}$, $\alpha=1,\ldots N_p$, are uniformly distributed in the interval $[0,2 \pi)$ and particle positions uniformly distributed over the spatial domain. 

\subsubsection{Steady state kinetic energy}

We monitor the time evolution of energy and observe that a steady state is reached with $(\Ri=90,\Sc=17)$ for $t V_s/h>10$, and with $(\Ri=90,\Sc=80)$ it takes a longer time $t V_s/h>400$. Qualitatively, this dependence on $\Sc$ can be understood by noting that an unbounded homogeneous Stokesian system destabilizes with a growth rate inversely proportional to the Schmidt number $\omega V_s/h \propto \Ri/\Sc$ \cite{pedley10a}. 

\begin{figure}[!h]
  \includegraphics*[width=0.45\linewidth]{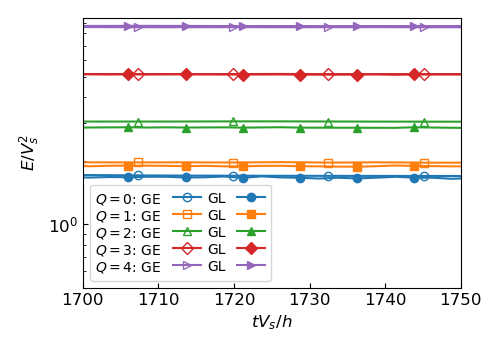}
  \includegraphics*[width=0.45\linewidth]{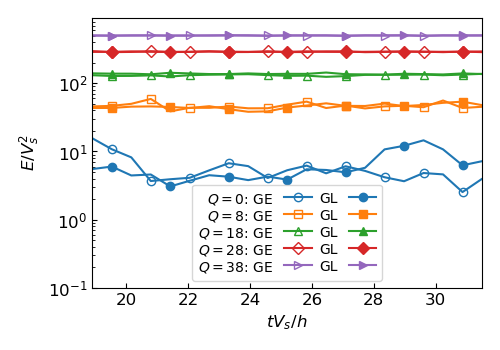}
  \caption{\label{fig:Eevol} Time evolution of the kinetic energy in the steady state for different values of flow rate $Q$ and with  $\Ri=90,\Sc=80$ (left), and $\Ri=90,\Sc=17$ (right).} 
\end{figure}

In \Fig{fig:Eevol}, we show the time evolution of energy with different flow rates $Q$ for $\Sc=17$ and $\Sc=80$ while maintaining $\Ri=90$ fixed. Note the excellent agreement between the results obtained from $\GE$ and $\GL$ equations. Next, we investigate the spatio-temporal structures for a range of flow rates.

\subsubsection{Zero flow rate: $Q=0$}
\begin{figure}
\centering
\includegraphics[width=0.35\linewidth]{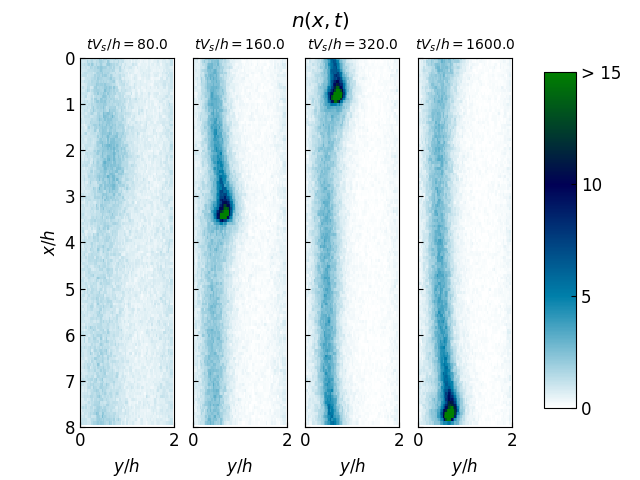}
\includegraphics[width=0.35\linewidth]{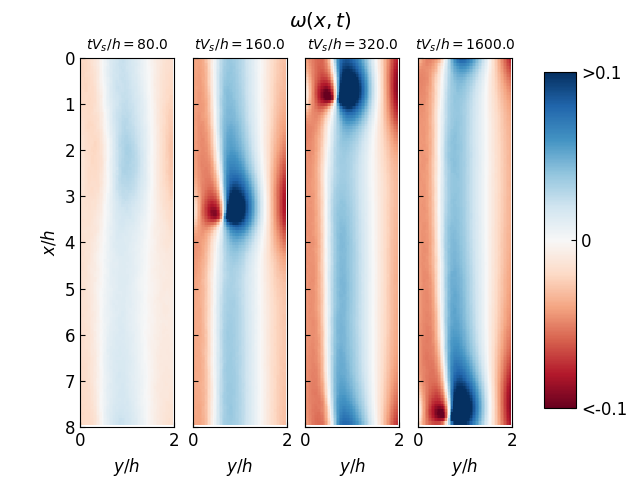}\\
\includegraphics[width=0.35\linewidth]{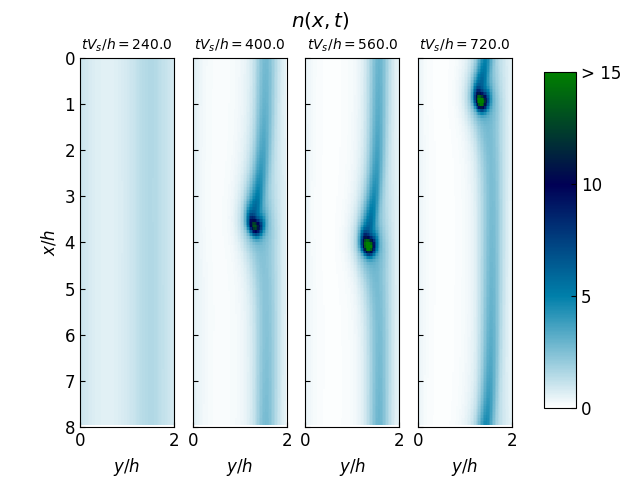}
\includegraphics[width=0.35\linewidth]{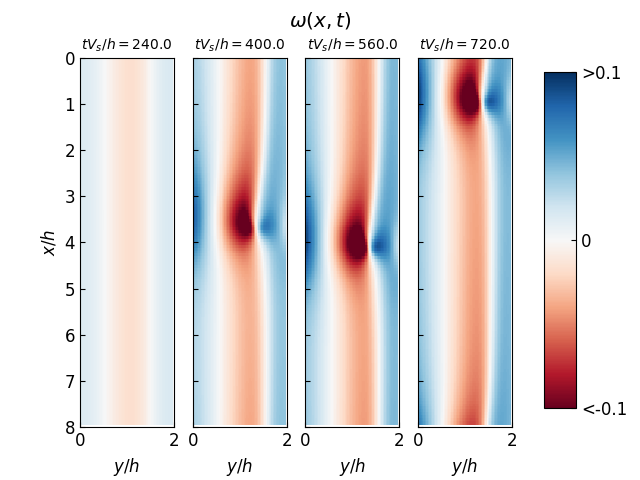}
\caption{\label{fig:Q0}  Pseudocolor plots showing the evolution of the cell number density and the vorticity  fields at different times for $Q=0,\Sc=80$. (Top panel) \GL formalism, and (bottom panel)  \GE formalism.}
\end{figure}

\begin{figure}
  \centering
  \includegraphics[width=0.35\linewidth]{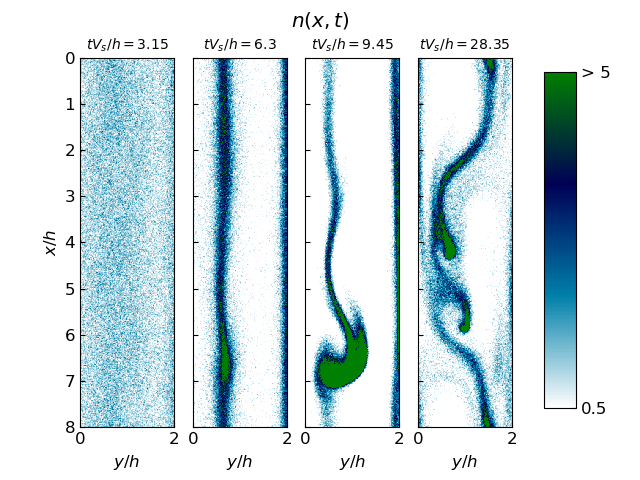}
  \includegraphics[width=0.35\linewidth]{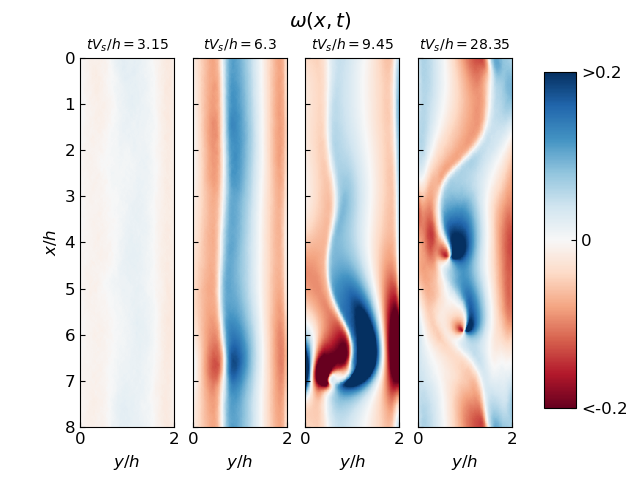}\\
  \includegraphics[width=0.35\linewidth]{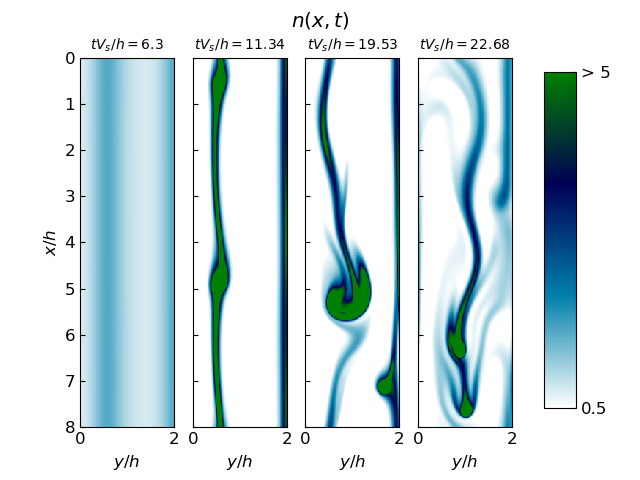}
  \includegraphics[width=0.35\linewidth]{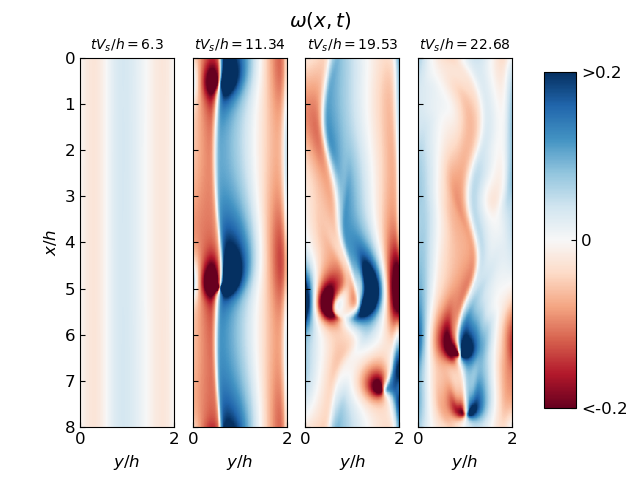}
  \caption{\label{fig:Q0_Sc17}  Pseudocolor plots showing the evolution of the cell number density and the vorticity fields at different times for $Q=0,\Sc=17$. (Top panel) \GL formalism, and (bottom panel)  \GE formalism.}
\end{figure}

We show the evolution of the cell number density and the vorticity snapshots at different times obtained using the \GL and \GE simulations for $\Sc=80$ in \Fig{fig:Q0}, and $\Sc=17$ in \Fig{fig:Q0_Sc17}. We note that the two methods give very similar solution evolution. For $\Sc=80$, the cells concentrate near one of the walls (depending on the initial perturbation), and we also observe varicose instability. The cells cluster and sink downwards (positive $x$-direction). The dynamics remain qualitatively similar for $\Sc=17$, and along with the formation of blip-like structures, we also observe large-scale meandering flows.

\subsubsection{Large positive $Q$}
\begin{figure}
\centering
\includegraphics[width=\linewidth]{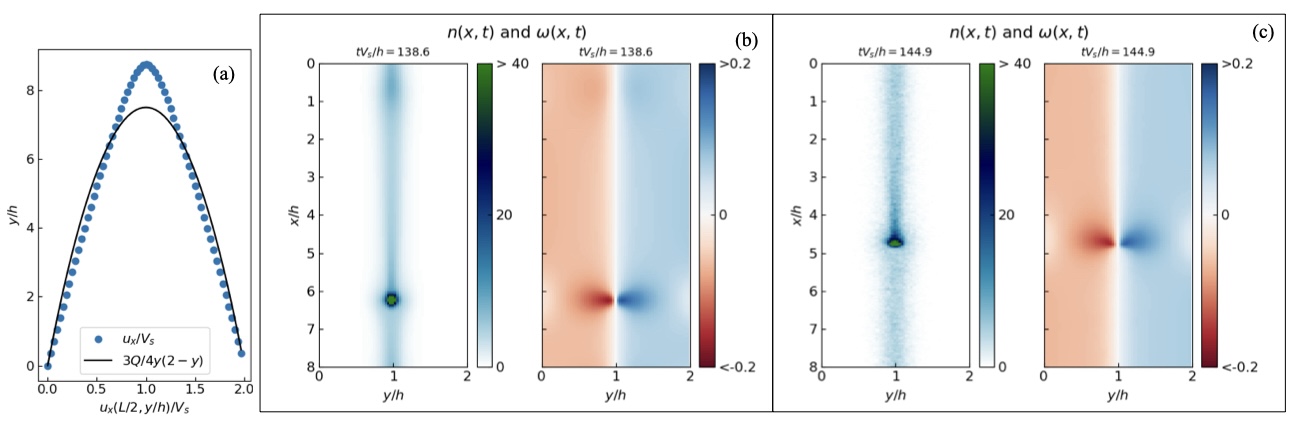}
\caption{\label{fig:Q1} (a) Average vertical velocity  $\int u_x(x,y) dx/L$ for $Q=10$.  Pseudocolor plots showing the steady state cell number density and the vorticity fields obtained by using (b) \GE formalism and (c) \GL formalism.}
\end{figure}

\begin{figure}
  \centering
  \includegraphics[width=\linewidth]{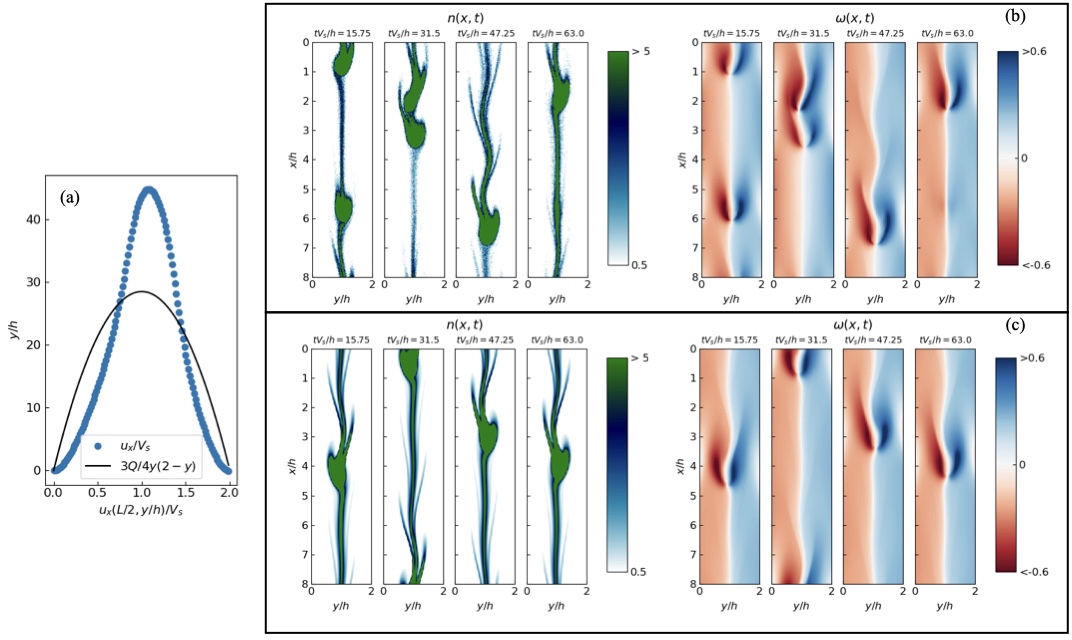}
  \caption{\label{fig:Q2} (a) Average vertical velocity  $\int u_x(x,y) dx/L$ at $x=L/2$ for $Q=38$. Pseudocolor plots show the evolution of the cell number density and the vorticity fields at different times for $Q=38$ using (b) \GL formalism and (c) \GE formalism.}
\end{figure}

For $Q=10, \Sc=80$ and $Q=38, \Sc=17$, the cells cluster around the centerline  ($y=h$) where the most significant departure from the Poiseuille profile is observed [see \Fig{fig:Q1}(a) and \Fig{fig:Q2}(a)].  In \Fig{fig:Q1} and \Fig{fig:Q2}, we show the snapshots of the cell number density and the vorticity field in the steady state and observe the formation of blips due to varicose instability. In both cases, we observe excellent agreement between Lagrangian and Eulerian formalism. Note that the analysis of \cite{hwang14a}, which only investigates streamwise perturbations, does not predict this instability.

Although not explicitly shown, the flow structures are similar to the $Q=0$ solution for small values and resemble the large $Q$ solution with increasing $Q$.

\subsubsection{Negative $Q$}
\begin{figure}
  \centering
  \includegraphics[width=0.45\linewidth]{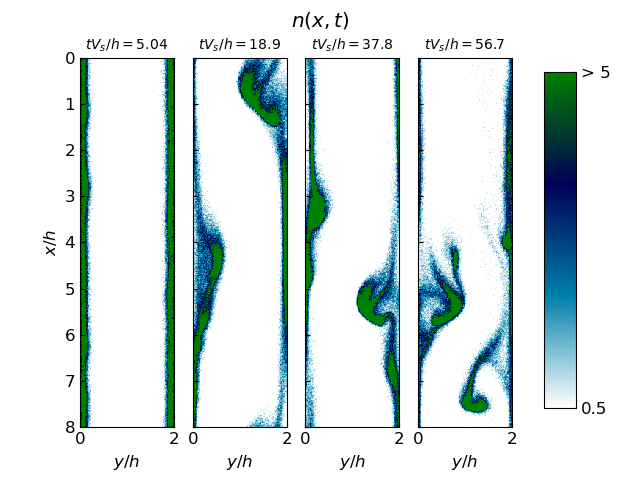}
  \includegraphics[width=0.45\linewidth]{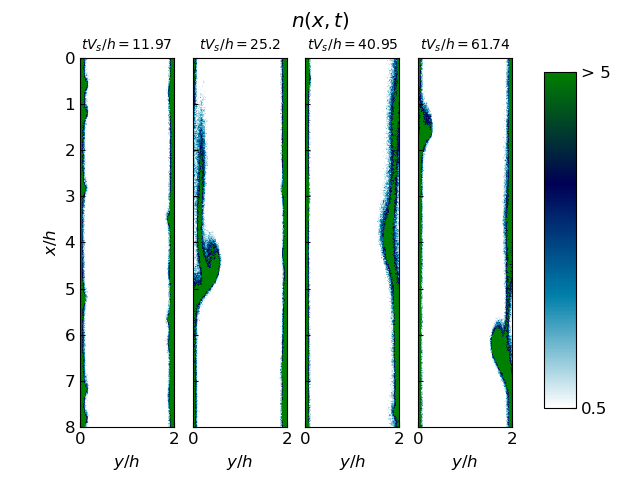}
  \caption{\label{fig:Qm24_Sc17}  Pseudocolor plots showing the evolution of the cell number density at different times for negative flow rates, $Q=-12$ (left panel) and $Q=-24$ (right panel), with $\Sc=17$ and the \GL formalism.}
\end{figure}

For negative $Q<0$, the flow and organism preferred swimming direction are the same, upwards. Therefore, in the absence of buoyancy effects the gyrotactic balance of gravitational and viscous torques would induce a mean biased swimming motion towards the walls. The plot in \Fig{fig:Qm24_Sc17} shows the time evolution of the cell concentration at different times for $Q=-12$ and $Q=-24$ with fixed $\Sc=17$ using the \GL approach. We observe that the cells accumulate close to the wall and start to form clumps, which at later times get peeled off from the wall due to an instability associated with their negative buoyancy and are pulled up by the flow as they move towards the center of the channel. The experiments of \cite{k85a} also observed a similar mechanism. Note that we do not present the solution obtained from the $\GE$ equations for $Q<0$ because for this range of parameters the numerical scheme is unable to conserve average cell density.

\section{Bioconvection in a Hele-Shaw cell \label{s:hshaw}}

In the above sections, we have explored the non-trivial plume structures that emerge for various positive and negative flow rates in a two-dimensional vertical channel.  As discussed earlier, a natural question arises as to whether we can effectively model confined bioconvection by making suitable approximations across the channel width.

\begin{figure}
  \centering
\includegraphics[width=0.5\linewidth]{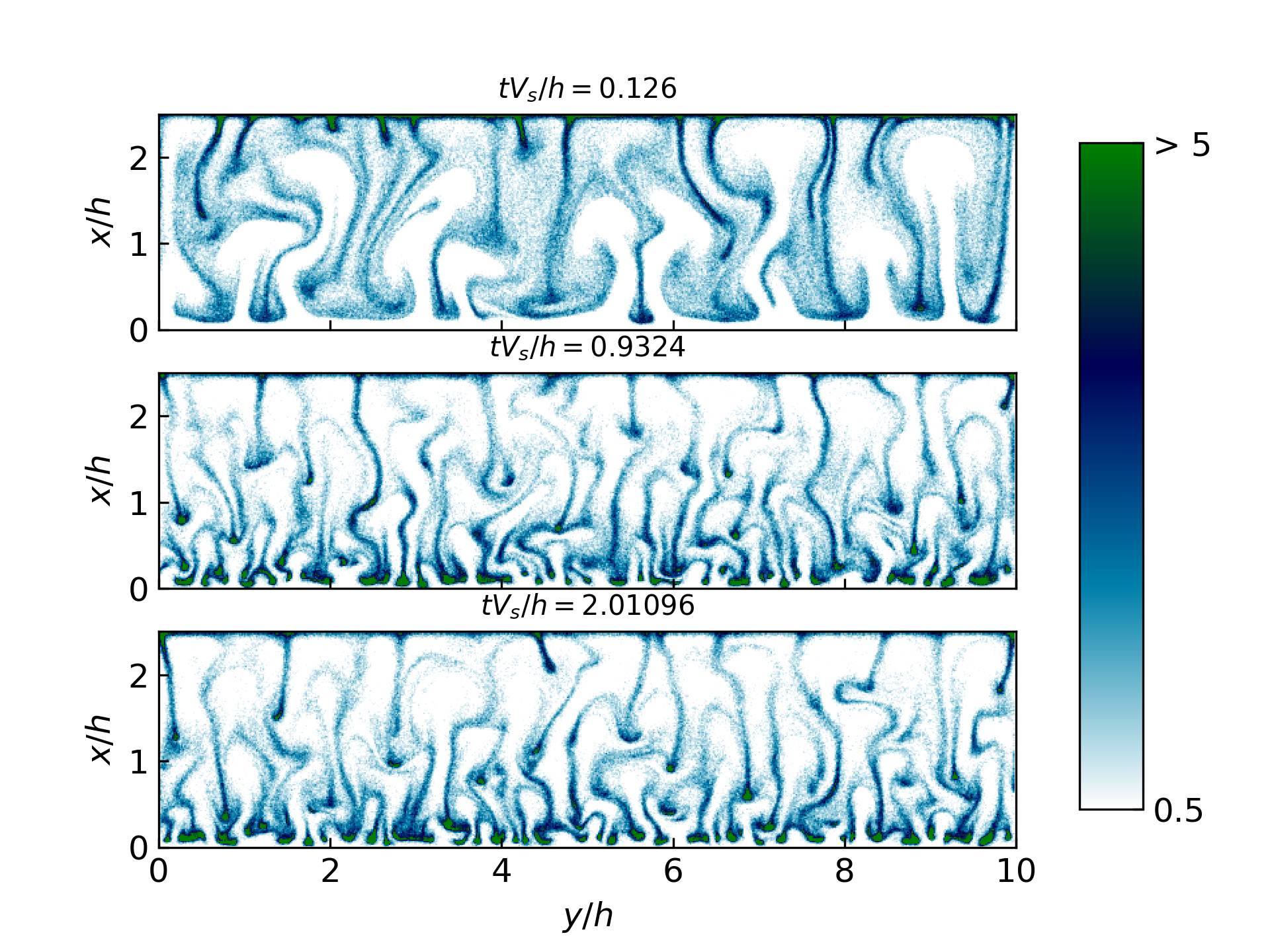}
\includegraphics[width=0.5\linewidth]{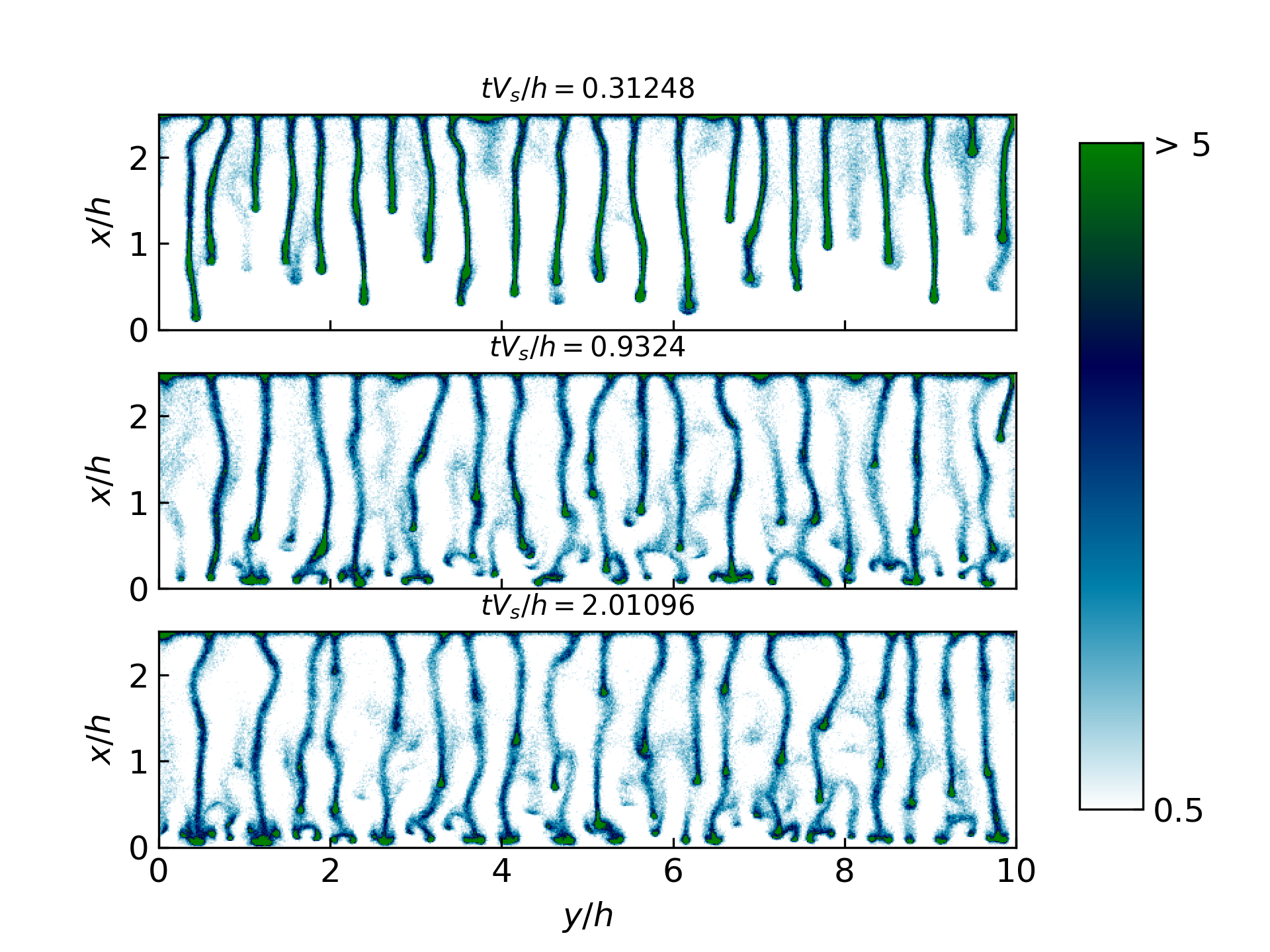}
\includegraphics[width=0.5\linewidth]{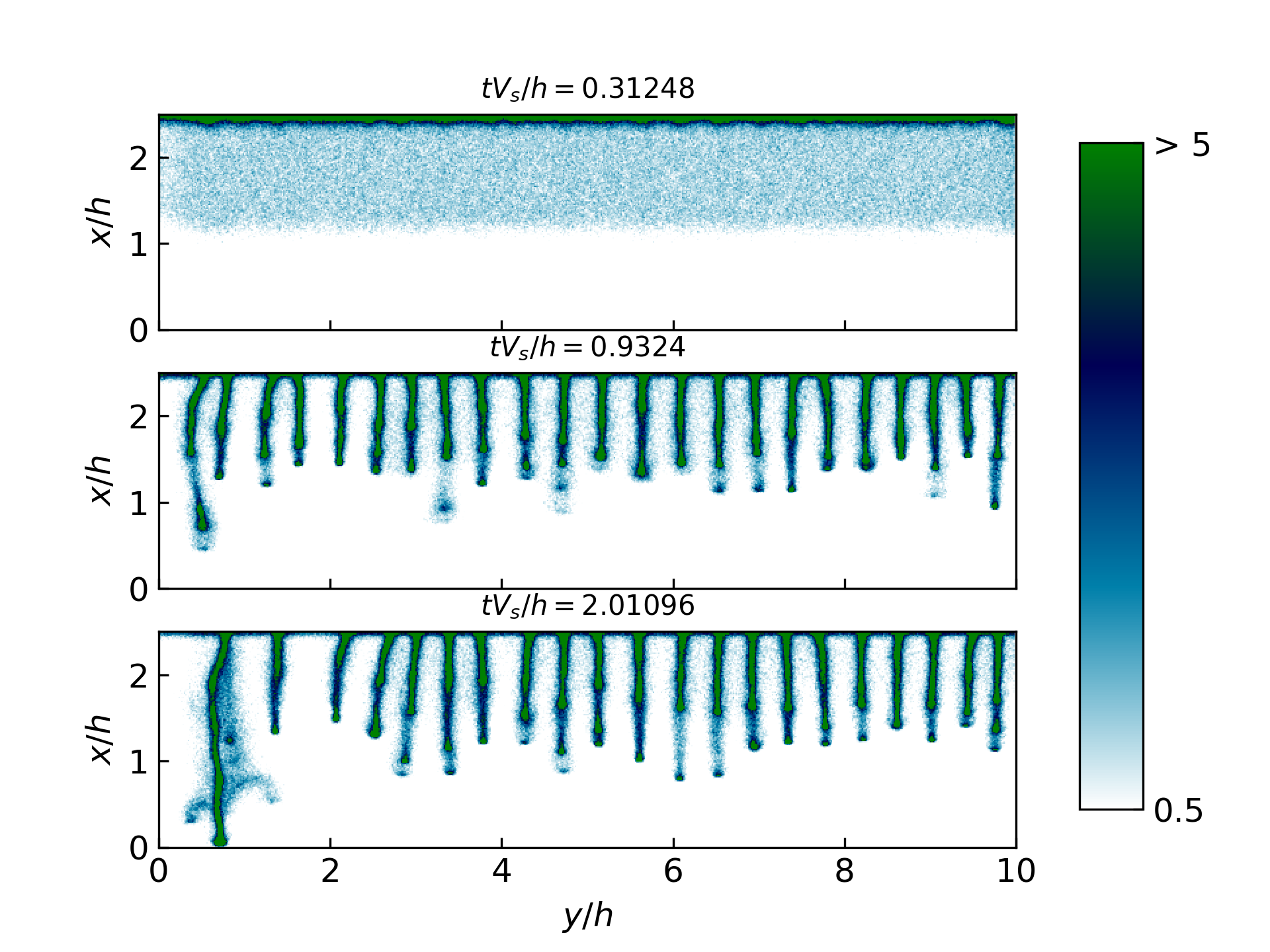}
\caption{\label{fig:HL} Pseudocolor plots showing the time evolution of the cell number density at different times in a Hele-Shaw setup for different values of $\Gamma=0$(top), $\Gamma=2.5$ (middle), and $\Gamma=12$ (bottom).}
\end{figure}

Therefore, we now extend our study to a quasi-two-dimensional setting by investigating gyrotaxis in a Hele-Shaw setup. To mimic experiments (see \subfig{fig:plumes}{b,c}; and \cite{bees2020advances}) we consider a rectangular domain of length $2h=10 \cm$, height $L=2.5 \cm$, and width $w=1 \mm$ inoculated with $n_0\sim 1.6 \times 10^6$ organisms (see \subfig{fig:sketch}{c}), this fixes $\Ri=2520$, and the other parameters are kept the same as in the previous sections. Since the confinement width is much smaller than the other two directions, we approximate its effect by adding a linear drag term $-\Gamma {\bm u}$ to \eqref{eq:ns}. Note that $\Gamma=12 {\mu}/w^2$ if a Poiseuille flow profile is assumed for the $z$-direction \citep{qui01}. Considering the variety of solutions discussed in the previous section, the validity of this linear drag approximation is indeed questionable. Nevertheless,  we study gyrotaxis in a vertical Hele-Shaw cell employing this two-dimensional setup with $\Gamma=0,2.5$ and $12$ (see \Fig{fig:HL}). For $\Gamma=0$, we observe that the gyrotaxis instability is initiated early on in the simulation, and the number density plots show a complex spatio-temporal structure similar to an earlier numerical investigation by \cite{hopkins02}. We note the stark contrast with experiments where regularly ordered bottom-standing plume structures are observed. In the presence of linear drag with $\Gamma=2.5$, we find regularly placed plumes similar to experiments. However, instead of forming bottom-standing plumes, we observe that plumes break into secondary structures when approaching the bottom wall. Further increasing $\Gamma=12$ increases the regularity of the structures, but now almost all of the plumes are initiated by an overturning instability at the top, not a gyrotactic instability throughout the Hele-Shaw cell, and they are unable to reach the bottom wall during the duration of the simulation.  Such a situation does arise in experiments, but for gradually tapered Hele-Shaw vessels rather than those with parallel walls (see \subfig{fig:plumes}{d}), suggesting that there is more occurring in the $z$-direction than anticipated.

Our simulations reveal that for intermediate values of drag, $\Gamma$, the concentration of organisms is similar but not identical to that observed in experiments; the dynamics near the bottom wall 
are very different. Furthermore, it is difficult to estimate the `best' value of $\Gamma$, to tune the simulations to get something similar.  This is likely due to the following three-dimensional effects.  
As the vessel is bounded at the top and bottom, the mean vertical flow rate is zero such that we are within the regime where a range of symmetric and asymmetric solutions exist between the two walls of the Hele-Shaw cell (i.e.~in the $z$-direction).  In particular, we expect asymmetric and meandering structures to dominate, such as in \Fig{fig:Q0_Sc17}.  Therefore, the normal assumption of Poiseuille flow within the Hele-Shaw cell is unlikely to be appropriate.
In addition, local upwelling flows are generated in at least half of the apparatus, and so cells become closely associated with both channel walls.  Thus they are not necessarily transported upwards by the flow in these regions, unless instabilities lead plumes of cells to peel off of the walls into the centre of the channel where they can be advected, such as in \Fig{fig:Qm24_Sc17}.

\section{Conclusions}
In this article we explore the role of confinement on the instabilities and dynamics of suspensions of gyrotactic microorganisms in two-dimensional channel flow and in Hele-Shaw cells, for parameters that are experimentally accessible. 
We tackle this problem using a combination of analytical techniques and two distinct numerical approaches, Eulerian and Lagrangian simulations.  
Remarkably, all three show excellent agreement, beyond our expectations.  
In particular, we find that asymmetric solutions are more stable than symmetric solutions for a realistic parameter range that we are not able to fully capture bioconvection and plume dynamics in a vertical Hele-Shaw cell with a two-dimensional approximation, strongly suggesting that commonly-adopted symmetry assumptions in confined geometries in the literature and their induced flow profiles are not appropriate.  

We start by exploring confined gyrotactic bioconvection in a channel as a function of the non-dimensional flow rate as well as the Schmidt, Richardson and P\'eclet numbers.  Using analytical approaches and numerical simulations for various flow rates, we have discovered the presence of asymmetric homogeneous solutions, that coexist with symmetric solutions. Even for the case of a zero mean flow rate, we find asymmetric flow profiles. Surprisingly, two independent numerical simulation methods (from Lagrangian and Eulerian perspectives) reveal that the asymmetric flow solution is in fact more stable than the symmetric profile that is often assumed and has been widely investigated \citep{hwang14}. 
Remarkably, all of the vertically homogeneous solutions closely obey a first integral, which we obtain by analysing the steady-state equations and confirm with simulations.

Next, we explore time-dependent solutions with numerical simulations in a vertical channel geometry and demonstrate the presence of complex spatio-temporal flow structures for a zero flow rate of $Q=0$. For $Q>0$, we verify the presence of a secondary varicose instability and the formation of travelling blips \citep{bees2020advances,hwang14,k85a}. For negative flow rates $Q<0$, we find that organisms migrate to the walls, as anticipated, but then this is followed by an instability that leads to structures that peel off from the walls and are transported upwards by the flow, later to migrate back to the walls. 

Finally, we extend our numerical analysis by studying gyrotaxis in a vertical Hele-Shaw cell by using a quasi-two-dimensional setting with a linear drag to mimic confinement. Although some of the results bear a resemblance to the experimental results it becomes necessary to attempt to fine-tune the simulations with the drag strength, $\Gamma$: not all features are correctly reproduced for a single value of $\Gamma$.  In essence, ignoring the migration of organisms towards the walls in locally upwelling flows and the complex, and typically asymmetric dynamics at slow flow rates leads to unsatisfactory results. 
Therefore, na\"ive averaging is not correct and we argue that it is essential to perform a full three-dimensional study to understand the effect of confinement in these active suspensions.  
Our study motivates future investigations of active matter in simple confined geometries, particularly at the dilute end of the spectrum.

%

\section*{Acknowledgement}
We offer our sincerest gratitude to John O.~Kessler, posthumously, for permission to use the images in Fig.~\ref{fig:plumes} from experiments joint with MAB.  John inspired a generation of researchers to explore gyrotactic bioconvection problems with his seminal studies of the phenomena in the 1980s \citep{k85a}. PP  acknowledge support from the Department of Atomic Energy (DAE), India under Project Identification No. RTI 4007, and DST (India) Project Nos. MTR/2022/000867,  and DST/NSM/R\&D HPC Applications/Extension/2023/08.


\bibliography{bibnew} 
\end{document}